\documentclass[sigconf,10pt]{acmart}

\AtBeginDocument{%
  }

\acmConference{MobiCom4AgeTech}{November 18--22, 2024}{Washington, D.C.}

\acmISBN{978-1-4503-XXXX-X/18/06}

\usepackage{array}
\usepackage{tabularx}
\usepackage{float}

\copyrightyear{2024}
\acmYear{2024}
\setcopyright{acmlicensed}\acmConference[ACM MobiCom '24]{The 30th Annual International Conference on Mobile Computing and Networking}{November 18--22, 2024}{Washington D.C., DC, USA}
\acmBooktitle{The 30th Annual International Conference on Mobile Computing and Networking (ACM MobiCom '24), November 18--22, 2024, Washington D.C., DC, USA}
\acmDOI{10.1145/3636534.3698118}
\acmISBN{979-8-4007-0489-5/24/11}

\begin{document}

\title[Enhancing EHR with Wearables: Monitoring Post-Surgical Symptoms in Seniors]{%
  Enhancing EHR Systems with data from wearables: An end-to-end Solution for monitoring post-Surgical Symptoms in older adults}

\author{Heng Sun}
\affiliation{%
  \institution{University of Florida}
  \city{Gainesville}
  \country{Florida}}
\email{hengsun@ufl.edu}

\author{Sai Manoj Jalam}
\affiliation{%
  \institution{University of Florida}
  \city{Gainesville}
  \country{Florida}}
\email{jalam.saimanoj@ufl.edu}

\author{Havish Kodali}
\affiliation{%
  \institution{University of Florida}
  \city{Gainesville}
  \country{Florida}}
\email{hkodali1@ufl.edu}

\author{Subhash Nerella}
\affiliation{%
  \institution{University of Florida}
  \city{Gainesville}
  \country{Florida}}
\email{subhashnerella@ufl.edu}

\author{Ruben D. Zapata}
\affiliation{%
  \institution{University of Florida}
  \city{Gainesville}
  \country{Florida}}
\email{rzapata@ufl.edu}

\author{Nicole Gravina}
\affiliation{%
  \institution{University of Florida}
  \city{Gainesville}
  \country{Florida}}
\email{ngravina@ufl.edu}

\author{Jessica Ray}
\affiliation{%
  \institution{University of Florida}
  \city{Gainesville}
  \country{Florida}}
\email{jessica.ray@ufl.edu}

\author{Erik C. Schmidt}
\affiliation{%
  \institution{University of Florida}
  \city{Gainesville}
  \country{Florida}}
\email{cavedivr@ufl.edu}

\author{Todd Matthew Manini}
\affiliation{%
  \institution{University of Florida}
  \city{Gainesville}
  \country{Florida}}
\email{tmanini@ufl.edu}

\author{Rashidi Parisa}
\affiliation{%
  \institution{University of Florida}
  \city{Gainesville}
  \country{Florida}}
\email{parisa.rashidi@bme.ufl.edu}

\renewcommand{\shortauthors}{Heng et al.}

\begin{abstract}
Mobile health (mHealth) apps have gained popularity over the past decade for patient health monitoring, yet their potential for timely intervention is underutilized due to limited integration with electronic health records (EHR) systems. Current EHR systems lack real-time monitoring capabilities for symptoms, medication adherence, physical and social functions, and community integration. Existing systems typically rely on static, in-clinic measures rather than dynamic, real-time patient data. This highlights the need for automated, scalable, and human-centered platforms to integrate patient-generated health data (PGHD) within EHR. Incorporating PGHD in a user-friendly format can enhance patient symptom surveillance, ultimately improving care management and post-surgical outcomes. To address this barrier, we have developed an mHealth platform, ROAMM-EHR, to capture real-time sensor data and Patient Reported Outcomes (PROs) using a smartwatch. The ROAMM-EHR platform can capture data from a consumer smartwatch, send captured data to a secure server, and display information within the Epic EHR system using a user-friendly interface, thus enabling healthcare providers to monitor post-surgical symptoms effectively.
\end{abstract}

\makeatletter
\def\@copyrightspace{\relax}
\makeatother

\settopmatter{printacmref=false}

\maketitle

\section{Introduction}
Over the past decade, mobile health (mHealth) apps have become increasingly popular for monitoring health conditions \cite{paglialonga_overview_2018}. While existing sensors and devices collect useful information for patients and researchers, such information is rarely shared with healthcare providers, potentially preventing timely interventions \cite{genes_smartphone_2018}. Existing provider portals lack real-time capabilities for monitoring patient symptoms, missed medications, physical function, social function, and community integration \cite{mohammed_real-time_2019}. These point to a critical unmet need for evidence-based decisions for utilizing automated, scalable, and human-centered platforms for integrating patient-generated health data (PGHD) within the electronic health record (EHR).  Regular surveillance of patient symptoms, activity and mobility patterns can facilitate better care management that results in favorable post-surgical outcomes. 

Technology that informs decision tools is a rapidly growing area in healthcare research \cite{tubaishat_perceived_2018, schnall_health_2018}.  However, decision tools are not typically provided with patient-derived or real-time information — they rely on in-clinic measures, known contraindications, and disease \cite{mohammed_real-time_2019}. There's a potential for close and ongoing patient monitoring to be highly valuable. For instance, timely detection of non-healing wounds can prevent severe complications \cite{han_chronic_2017}. Yet, healthcare providers solely rely on snapshots of patient recall outside of the hospital setting \cite{rothman_use_2009}. For those who have adopted using PGHD, data are often limited to either symptoms or sensor data, with almost none collecting both simultaneously \cite{austin_use_2020}. Synchronous data streams of symptoms and sensor data have the potential to contribute to providers’ decision-making \cite{islind_shift_2019, nerella_ai-enhanced_2024} to foster a deeper and more accurate understanding of a patient’s recovery or lack-there of \cite{cohen_integrating_2016}.

Despite the recent development of integrated mHealth interfaces for EHR systems, integrating PGHD within the already overly complex clinical workflow still poses many usability and integration challenges \cite{genes_smartphone_2018}. In 2019, Austin et al. \cite{austin_use_2020} examined the use of PGHD in clinical care by providing a survey to current and potential users of PGHD. Participants showed how PGHD can be used in targeted areas of healthcare system, and most participants cited the potential for PGHD to enhance care delivery and outcomes, but also indicated substantial barriers to more widespread PGHD adoption across healthcare systems. In 2023, Webber et al. \cite{webber_integrating_2023} provided a practical guide for integrating PGHD into EHRs, highlighting the importance of stakeholder engagement, characteristics of high data quality, and PGHD in practice in patient-driven registries. Beyond these issues, there is a lack of usability and workflow guidelines to facilitate patient sign-up and providers’ processing of incoming PGHD. Also, sensors from smartwatches generate a vast quantity of data; they require meaningful summarization and visualization that is integrating and informative for clinical practitioners \cite{robert_m-health_2006}. Additionally, the existing systems are severely limited in context with many only integrating a single measurement (e.g. glucose monitor data) into EHR \cite{kumar_automated_2016}.  Other work is done in simulated environments \cite{marceglia_standards-based_2015} using research platforms such as the Informatics for Integrating Biology and The Bedside (i2b2) instead of an actual EHR system \cite{pfiffner_c3-pro_2016}, or without an end-to-end solution \cite{park_development_2016,leijdekkers_improving_2015}.

In this study, we develop an integrated EHR extension specifically designed to address challenges faced by diverse patient populations, with a particular focus on older adults. Given that age-related factors such as delayed wound healing, increased fall risk, and mobility issues significantly impact post-surgical care, the extension provides healthcare providers with real-time monitoring capabilities by importing and displaying data from pre-configured smartwatches. Data collected from the smartwatches is transmitted and displayed on the interface instantaneously. Additionally, this extension includes a notification system that alerts healthcare providers to any urgent alerts from patients, thereby enhancing care delivery and outcomes.

\section{Architecture}
The ROAMM-EHR platform functions as a comprehensive system for managing patient data. It consists of three key components: a smartwatch app, an Amazon Web Services (AWS) server, and an Epic app (as illustrated in Figure 1). The smartwatch app acts as an initial data collection point, collecting information directly from patients. The collected data are then encrypted and securely transmitted to the server. The server acts as the central hub, receiving and analyzing patient data from the smartwatch app. Additionally, the server provides essential Application Programming Interfaces (APIs) for the Epic app to seamlessly retrieve the analyzed data from it. Finally, the Epic app serves as the user interface for healthcare providers, displaying the collected data in a clear and concise format. This holistic approach streamlines the entire patient data collection and analysis process, ultimately improving healthcare providers' efficiency and decision-making.

\begin{figure}[H]
    \centering
    \includegraphics[width=1\linewidth]{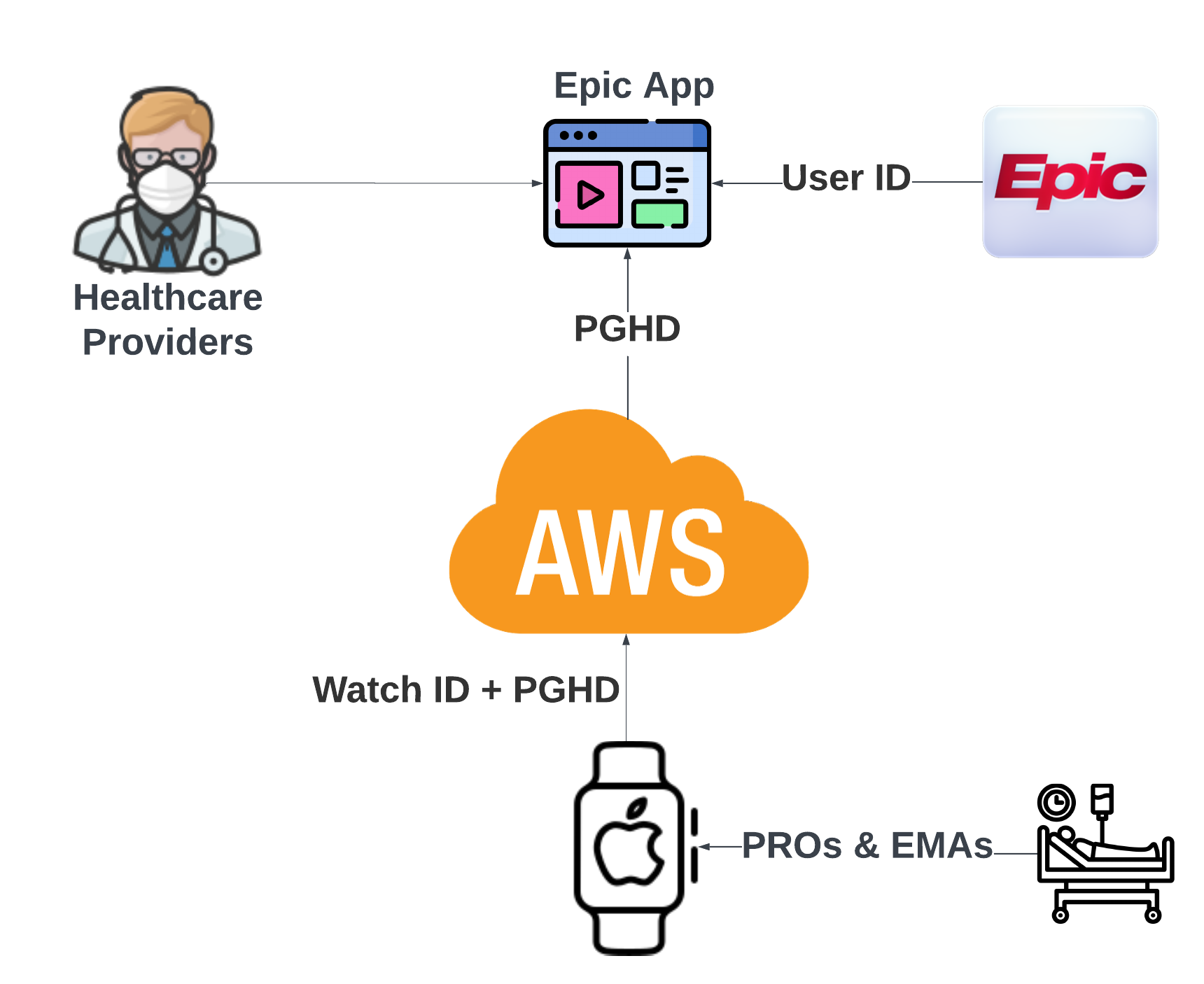}
    \caption{High-level architecture of ROAMM-EHR platform. It comprises three main components: the smartwatch app collects PROs and EMAs and sends data to the AWS server; the AWS server analyzes and forwards this data to the Epic app; and the Epic app displays the processed data to healthcare providers based on user ID decrypted from the Epic server.}
    \label{fig:enter-label}
\end{figure}

\subsection{Smartwatch App}
The smartwatch application was designed and developed for the Apple Watch using Swift version 5 and Swift UI framework \cite{karnati_roamm_2021}. The architecture for the application follows the View Model design pattern. The app gathers data from multiple sensors, including accelerometer, gyroscope, GPS, heart rate monitor, and UV exposure sensor, each generating data in diverse formats and resolutions. Additionally, the application features an interactive watch UI that prompts users in sessions to gather Patient-Reported Outcomes (PROs) \cite{weldring_patient-reported_2013} and Ecological Momentary Assessments (EMAs). These sessions present questions based on the date and time and allow users to choose their answers directly on the watch. Figure 2 shows various examples of these PROs prompts on an Apple Watch. The PROs collected include physical symptoms such as pain levels, mobility, and wound healing progress. The EMAs consist of prompts related to daily activity, mood, and cognitive function.

\begin{figure}[H]
    \centering
    \includegraphics[width=1\linewidth]{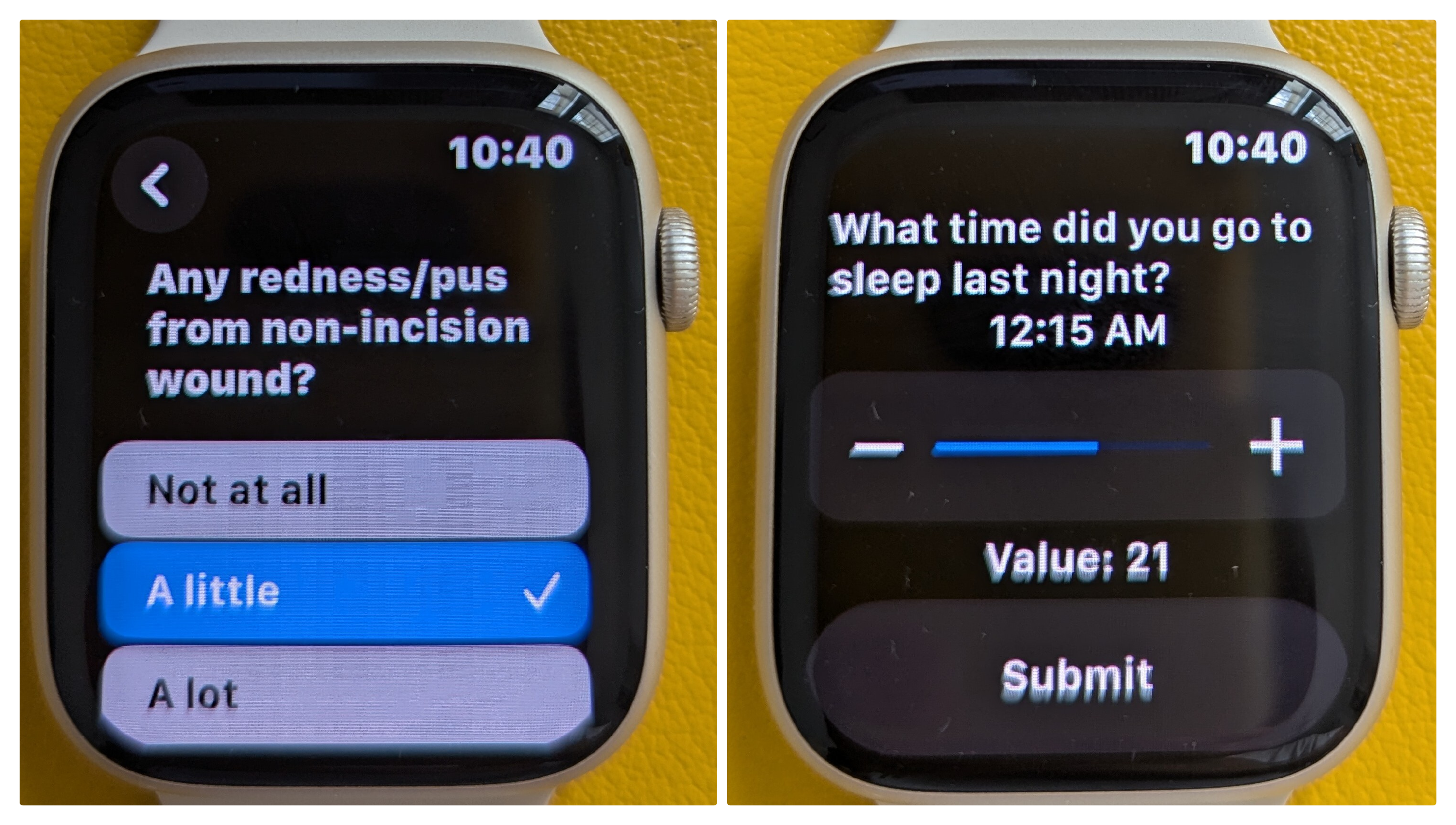}
    \caption{Examples of PROs in Apple watches.}
    \label{fig:enter-label}
\end{figure}

The application also facilitates cognitive assessments, meticulously capturing response times. All data collected are securely transferred to the server for further processing and analysis. To provide a customized user experience, the application adapts its features based on configurations (campaigns) dynamically downloaded from the server. These configurations define various parameters for data collection, such as sensor activation/deactivation, data collection frequencies, raw data aggregation intervals, data feature summarization (e.g., mean vector magnitude), relevant patient-reported outcomes, and selected cognitive tests. This dynamic adjustment capability ensures that the application provides a tailored and responsive data collection experience, optimized for the individual needs of each user.

\subsection{Server}
The ROAMM-EHR platform, supported by a server infrastructure illustrated in Figure 3, facilitates a wide range of functions, including managing authentication methods, ensuring data privacy across campaigns, interacting concurrently with multiple smartwatches, configuring and receiving data, storing them in a centralized database, and supporting web-based user interfaces. Leveraging a cloud-based architecture offers significant benefits such as flexibility, adaptability to changing resource requirements, high availability, and robust security. AWS provides multiple layers to safeguard data and infrastructure. One of the key services we leveraged in the ROAMM-EHR platform is AWS Lambda, which runs small pieces of code in response to events. This allows the platform to process patient data quickly without needing to manage complex server setups. Lambda functions are protected by JSON Web Tokens (JWT) token-based authentication, ensuring secure API access, and are hosted within a Virtual Private Cloud (VPC) to isolate computing resources for data security and integrity. Additional services like AWS Elastic Compute Cloud (EC2) and Relational Database Service (RDS) enforce strict access controls, using tools like SSH key-pairs and Identity and Access Management (IAM) policies to manage permissions and regulate traffic. Only authorized personnel can modify data or configurations. This comprehensive, multi-faceted approach ensures robust cloud security across all aspects of the platform.

\begin{figure}[H]
    \centering
    \includegraphics[width=1\linewidth]{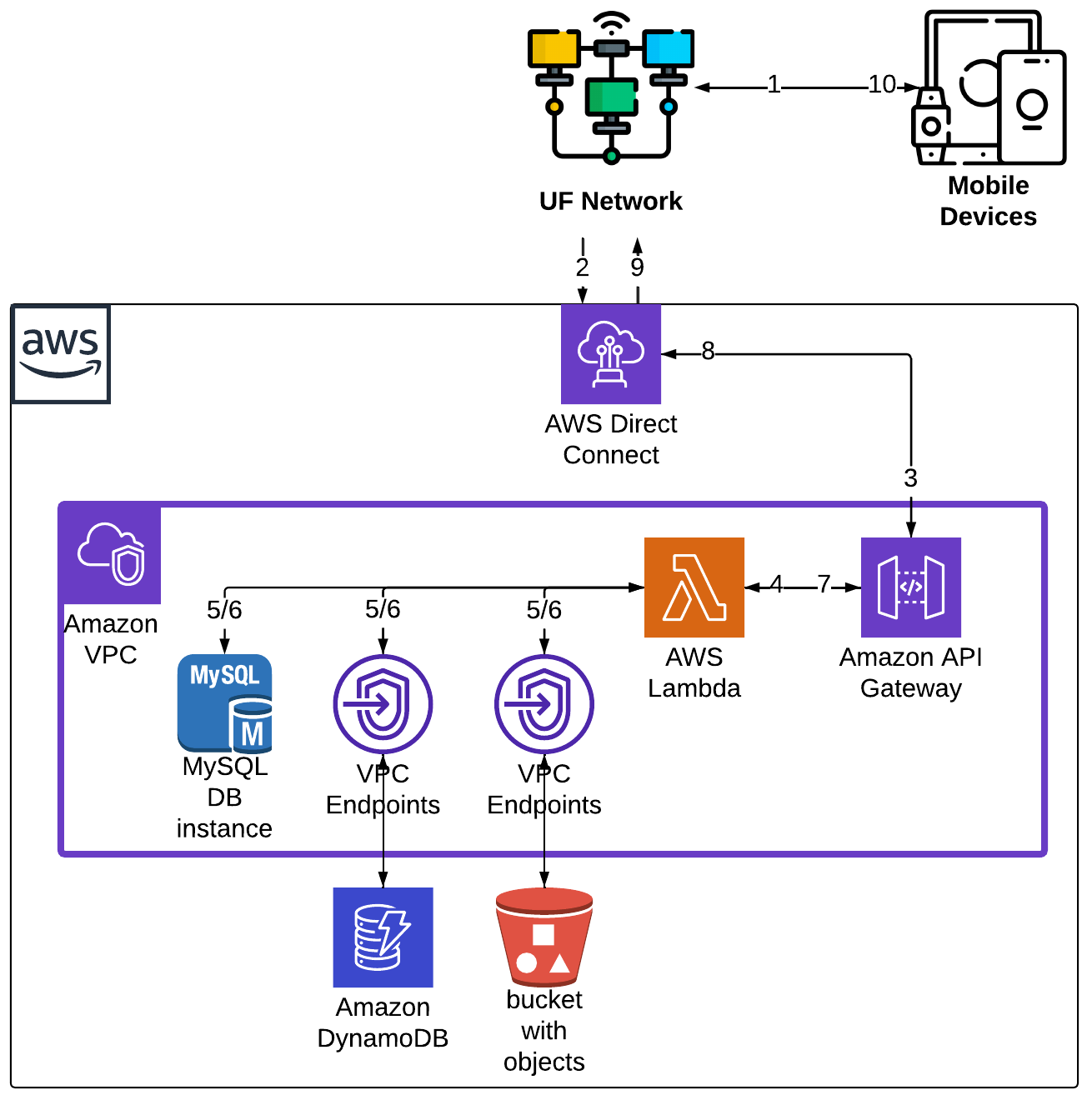}
    \caption{The server architecture involves mobile devices connecting to the AWS API Gateway via the University of Florida (UF) network and AWS Direct Connect. Data is fetched through a Lambda function, which retrieves information from the database. The processed data is then sent back to the mobile devices.}
    \label{fig:enter-label}
\end{figure}
 
The ROAMM-EHR platform's implementation utilizes AWS as the main cloud server to receive, analyze, and transmit data. AWS Lambda, a serverless computing service that runs code in response to events and automatically manages the underlying computing resources, is used extensively to handle data exchanges. Besides, it's also utilized to connect to AWS API Gateway to create different APIs for the smartwatch app and the Epic app, using Python as the default language for these Lambda functions. These APIs include a patient registration API, which allows for the creation of new patient records; a decryption API, which decrypts parameters in links sent from the Epic server; a clinical API, which retrieves individual patient clinical data for visualization purposes; a patient list API, which provides a list of all patients' basic information; and an authentication API, which verifies user credentials.

\begin{table}[ht]
\caption{The trigger value for notification detection system}
\centering
\begin{tabularx}{\columnwidth}{|>{\centering\arraybackslash}X|>{\centering\arraybackslash}X|}
\hline
 Watch Inquiry Questions & Threshold Trigger Value for Email Notifications\\
\hline
Redness spreading away from surgical incision?& A lot \\
\hline
Edges of surgical incision separated/gaped open? &A lot \\
\hline
Do you have any redness or pus leakage coming from a non-incision wound?&A little \\
\hline
Did you have changes in the appearance of your non-incision wounds?&A lot \\
\hline
Fever or raised temperature?&yes \\
\hline
Difficulty walking a block or 250 feet?&Unable/need help \\
\hline
Did you have a new fall but DID NOT cause injury&yes \\
\hline
Did you have a new fall but DID cause injury?&yes \\
\hline
Current incision pain?&>=8 \\
\hline
Maximal incision pain experienced today?&>=8 \\
\hline
Today, did you have new pain at rest in the foot on surgical limb?&>=4 \\
\hline
\end{tabularx}
\end{table}

For data storage, Amazon DynamoDB is utilized, managed by MySQL, to ensure efficient and scalable data management. All data collected from the watch is stored in DynamoDB. 

Additionally, a detection system is implemented to monitor all PROs metric received from the watch. If the PROs metric exceeds a certain threshold identified listed in Table 1, the server sends a notification to the Epic app and an email to healthcare providers, informing them of the situation.

\subsection{Epic App}
The frontend of the ROAMM-EHR platform was developed to ensure responsiveness, performance, and user-friendliness. The primary technologies and frameworks used include React.js, a JavaScript library for building user interfaces, chosen for its component-based architecture and efficient rendering capabilities. Material UI, a popular React UI framework, provided pre-designed components and styles, ensuring a consistent and visually appealing design. Chart.js, a JavaScript library, was used for creating responsive and interactive charts, ideal for visualizing healthcare data. Jotai, a state management library, simplified the management of complex states in the application. Axios, a promise-based HTTP client, facilitated API requests, chosen for its simplicity and ease of use. Additionally, Day.js, a lightweight JavaScript library for date manipulation, ensured accurate and efficient handling of dates and times. By utilizing these technologies, it streamlines the development process, enhances scalability for future modifications and incorporates new features swiftly.

\begin{figure}[H]
    \centering
    \includegraphics[width=1\linewidth]{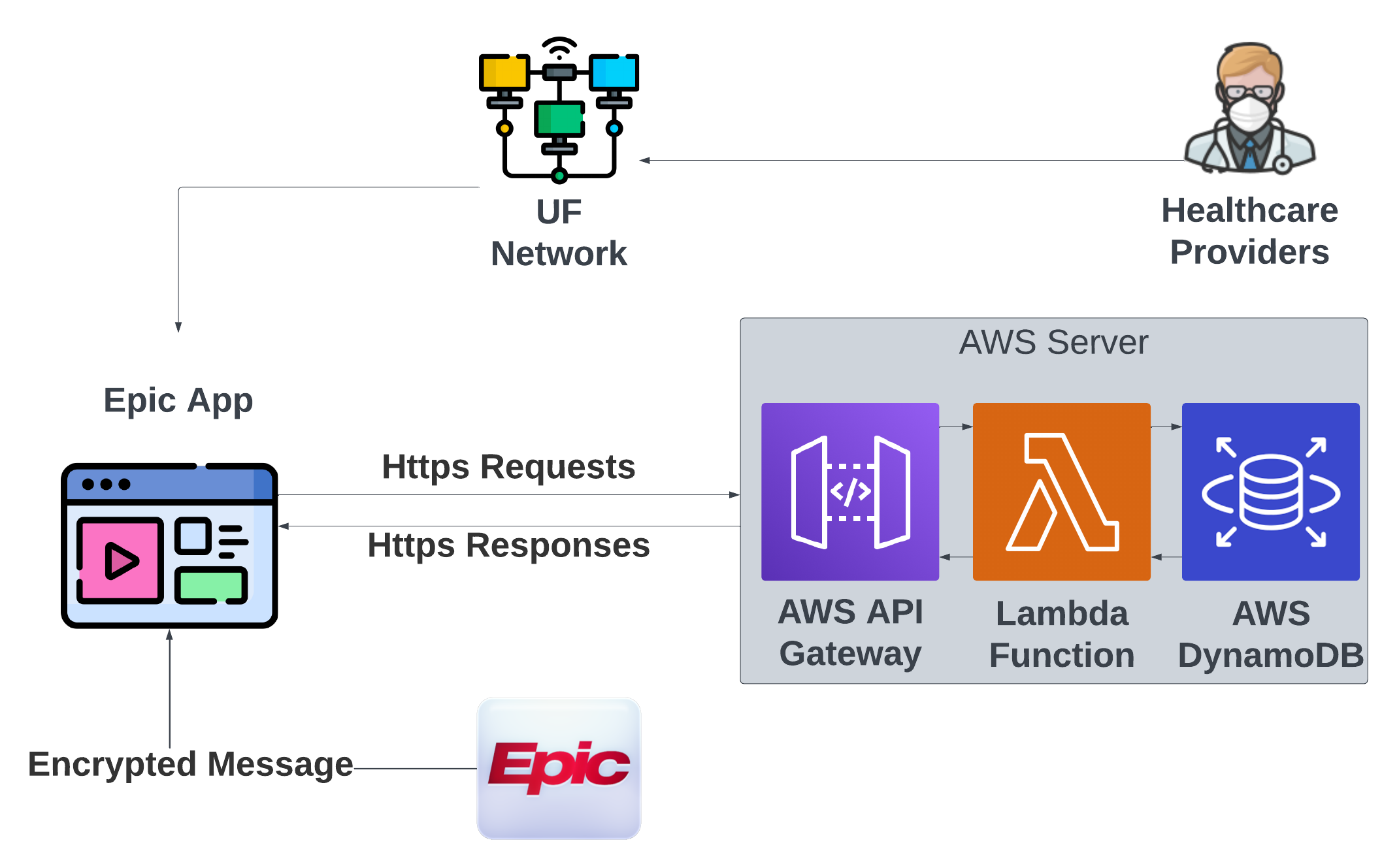}
    \caption{Architecture of the Epic app. The healthcare provider is asked to connect to UF VPN to access the Epic app. Upon accessing the Epic app, the Epic server sends the encrypted user ID to the app. The app interacts with the AWS server via https requests, processed by the AWS API Gateway and Lambda functions for data retrieval and filtering from the database. Responses are then sent back to the app through the same secure channel and displayed on the relevant page.}
    \label{fig:enter-label}
\end{figure}
\begin{figure*}
    \centering
    \includegraphics[width=1\linewidth]{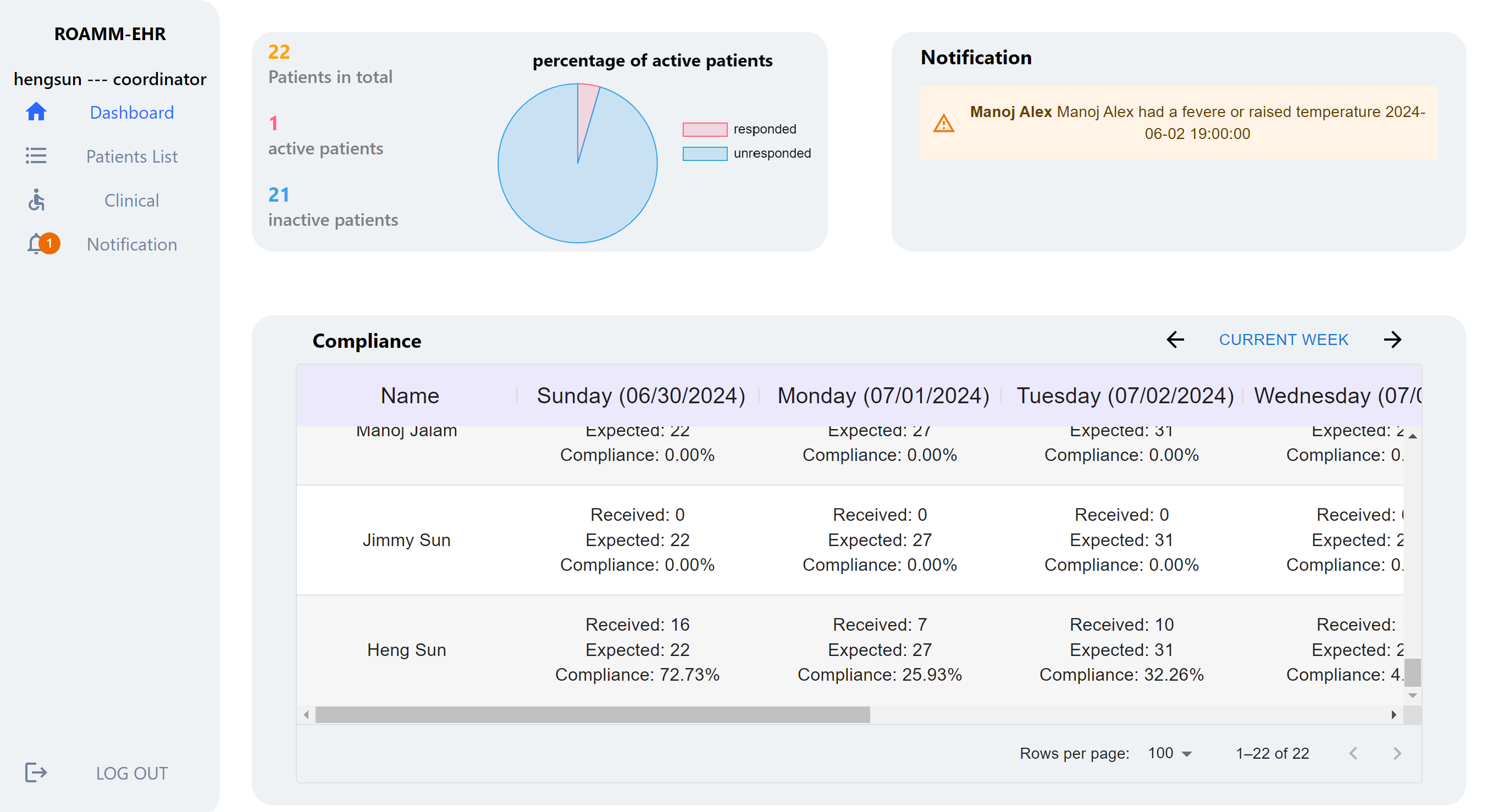}
    \caption{Overview of patients’ information. It features a pie chart displaying the percentage of responded patients, a notification box highlighting important updates and alerts, and a compliance table showing the compliance rates of all patients over the most recent week.}
    \label{fig:enter-label}
\end{figure*}

Figure 4 depicts the architecture of the Epic app. The application access is restricted to UF health network. Healthcare providers establish a UF VPN connection to access the Epic app. Upon accessing the Epic app, the Epic server sends the user ID to the app. Based on the user ID, the app identifies the role as either a provider or a coordinator. Only coordinators can use two specific features: registering a new patient and editing an existing patient. Within the app, user interactions trigger https requests, processed by the AWS API Gateway and Lambda functions. These functions retrieve and filter data from the database. Finally, processed data is delivered back to the Epic app through the same secure channel for display on relevant pages. This architecture ensures the confidentiality and integrity of patient data throughout the access and retrieval process.

The Epic app user interface was designed with a focus on usability and accessibility. Key considerations included responsive design, ensuring the website is accessible on various devices, including desktops, tablets, and smartphones. Intuitive navigation was implemented to simplify access to different sections of the website, such as the dashboard, patient list, clinical data, notifications, login, and patient registration. User-centric features were incorporated to cater to the needs of healthcare providers, such as real-time data visualization, easy patient management, and quick access to critical alerts. 

The Epic app offers a range of features designed to support healthcare providers in monitoring patient data effectively. The key features and functionalities of the web application are described below.

\subsubsection{Overview of Patients' Information}
The Epic app offers an overview of the healthcare provider's account and displays patient data for the most recent week (as illustrated in Figure 5), with an option to select different weeks. A pie chart created using Chart.js represents the percentage of active patients (those who answered prompts) versus inactive patients (those who did not answer prompts). Additionally, a short version of the notification block, which shows the latest  severe clinical status notifications from patients, is included. Finally, the compliance table at the bottom part of the dashboard summarizes the most recent week's expected replies from patients, the received replies, and their compliance rates in a table created by using data grid library developed by Material UI.

 \subsubsection{Categorical Responses Data Visualization}
The Epic app also displays data received from the watch, categorized into two types: discrete and numeric. Each type is visualized with appropriate charts and graphs, as shown in Figure 6. The PGHD is collected and analyzed on the server based on the date and the selected question, allowing for detailed and specific data insights. Different dates and questions display their corresponding values, enabling comprehensive tracking of patient health metrics over time. In the clinical page, the first question section is interactive and clickable. This feature allows healthcare providers to select different questions within this section, facilitating easy comparison and analysis of various data points.

For numeric data visualization, the Chart.js library is used to create a line chart, with the X-axis representing time and the Y-axis representing values. Numeric data encompasses sensor data such as heart rate, step count, distance traveled, remaining battery life, as well as the numeric responses to questions from the watch.

\begin{figure}[H]
    \centering
    \includegraphics[width=1\linewidth]{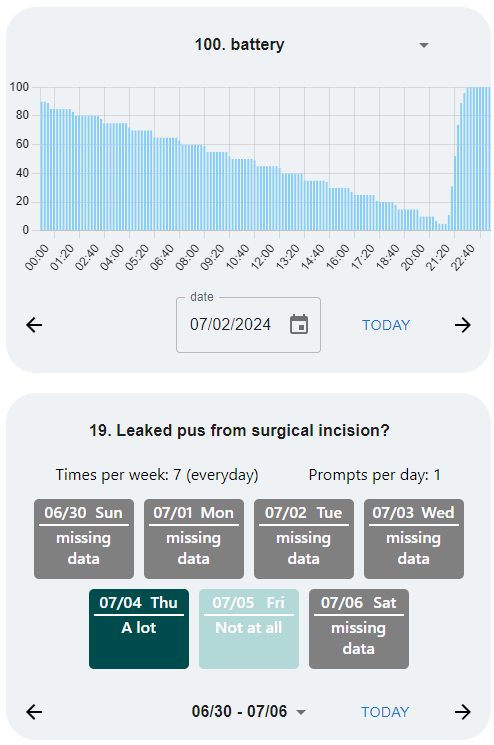}
    \caption{Categorical Responses Data Visualization. For numerical data, the top section displays a line chart, while the bottom section shows a one-week calendar for discrete data.}
    \label{fig:enter-label}
\end{figure}

\begin{table}[ht]
\caption{Color codes and meanings}
\centering
\begin{tabularx}{\columnwidth}{|>{\centering\arraybackslash}X|>{\centering\arraybackslash}X|}
\hline
 Colors & Description\\
\hline
Gray & Indicates missing data if date has passed \\
\hline
Yellow & Shows pending entries if date hasn't been reached \\
\hline
Neutral & Used for received data, selecting first unused color from predefined set \\
\hline
Black & No prompt asked on that date \\
\hline
\end{tabularx}
\end{table}

For discrete data, a one-week calendar visually represents the values retrieved from the watch. The calendar blocks change color based on the data received (Table 2): gray for missing data if the date has passed, yellow for pending entries if the date hasn't been reached, black if prompt was not asked on that date and a unique neutral color from a predefined color set for received data. To better serve healthcare providers, the interface also offers a download feature for them to selectively download specific datasets, including clinical prompts data, GPS data, sensor data, and health kit data.

\subsubsection{Notification}
The Epic app incorporates a feature that alerts users to critical events and updates to ensure timely responses to significant changes in patient conditions. This feature employs a color-coded system to differentiate message types: green denotes positive news (patient recovery), blue represents regular information (system updates, successful report reception, etc.), orange signifies severe situations (conditions from Table 1), and red indicates emergency information (conditions from Table 1 persisting for several days).

\section{Discussion}
Prior to developing the ROAMM-EHR platform, a pilot deployment was conducted to assess its feasibility and accessibility. This pilot initiative involved a select group of healthcare providers who utilized AI-generated test data to evaluate the platform's capabilities. The positive outcomes from this pilot phase, including the successful identification and timely intervention for complications, validated the platform's potential and informed our decision to develop the current version of ROAMM-EHR platform.

The ROAMM-EHR platform addresses critical gaps in current healthcare technology, particularly in integrating real-time PGHD with EHR \cite{paglialonga_overview_2018}. This integration is crucial for timely and effective clinical decision-making, especially in scenarios such as postoperative patient monitoring. It offers several advantages. Firstly, it seamlessly integrates PGHD into existing EHR systems, capturing data from consumer smartwatches and integrating it with platforms like Epic EHR. This capability enables real-time monitoring and enhances clinical decision-making by providing continuous surveillance of patient symptoms and activities. Healthcare providers can leverage this data to make more informed and timely decisions, potentially improving patient outcomes through proactive intervention.

Additionally, the platform incorporates an auto-notification system that ensures swift responses to severe situations by automatically sending alerts. This feature not only enhances patient safety but also maintains consistent communication among care teams, mitigating risks associated with delays in critical information dissemination. Moreover, the use of distinct colors for various types of notifications allows healthcare providers to quickly prioritize responses based on the urgency indicated. This visual differentiation aids in efficient workflow management and supports timely interventions, thereby optimizing patient care delivery. Designed on the AWS infrastructure, the platform ensures scalability, high availability, and robust security, essential for maintaining data integrity and complying with stringent healthcare regulations. Furthermore, the user interface is meticulously designed to meet the specific needs of healthcare providers, prioritizing usability and accessibility. This design focus facilitates easier adoption and seamless integration into existing clinical workflows, enhancing overall operational efficiency.

Physician feedback played a critical role in shaping the ROAMM-EHR platform. Early-stage user testing with physicians revealed challenges related to data overload and integration with clinical workflows. To address these, we simplified the user interface and introduced an alert prioritization system. Adoption of the platform has been encouraged through ongoing physician education and by emphasizing its potential to reduce post-surgical complications through real-time monitoring.

The ROAMM-EHR platform holds the potential in transforming how PGHD is integrated and utilized within EHRs, particularly in enhancing patient care and outcomes in post-surgical settings. However, it is important to acknowledge certain limitations. Despite its user-friendly design, healthcare providers may face initial challenges in adapting to new interfaces and integrating this technology into their daily practices \cite{clarke_how_2016}. Furthermore, the limited battery life of smartwatches can hinder continuous monitoring, requiring frequent recharging, which may disrupt the monitoring process. Another limitation is the upfront costs associated with implementing the monitoring system, including the procurement of smartwatches, backend infrastructure setup, and application development, which may present financial barriers for healthcare organizations. 

While smartwatches offer convenient real-time data collection, they may present accessibility challenges for older adults, particularly those with vision impairments. In the future, we plan to develop a companion smartphone and desktop application to ensure that users can complete EMAs more comfortably on devices with larger screens. Besides, our efforts will also be directed towards optimizing user experience, streamlining interface tutorials, and enhancing the transmission and response speeds of both the smartwatch and the AWS server. These initiatives aim to improve accessibility, usability and learning efficiency, as well as to ensure a more responsive and reliable system for healthcare providers across a more diverse patient populations. Addressing these challenges through ongoing training, support, and cost-effective deployment strategies will be crucial for maximizing the benefits of the ROAMM-EHR platform and ensuring its successful integration into routine clinical practice.

\section{Conclusion}

The ROAMM-EHR platform represents a significant step forward in integrating PGHD with EHRs, addressing a critical gap in real-time patient monitoring. By integrating consumer smartwatches, a robust cloud-based server, and a user-friendly interface, the platform provides healthcare providers with continuous, real-time data on patient symptoms and activities. This integration allows for more informed and timely clinical decisions, particularly in post-surgical care, where early intervention can prevent complications and improve patient outcomes. The platform's user-centric design ensures that healthcare providers can easily navigate and utilize the data, enhancing its adoption and effectiveness in clinical settings. The system's automatic notification feature further enhances responsiveness, ensuring that healthcare providers receive immediate alerts when patient conditions warrant attention, thereby improving overall patient outcomes. 

\section{Acknowledgements}
This study is supported by the National Institutes of Health’s National Institute on Aging (NIH/NIA), granted to the Department of Biomedical Engineering at the University of Florida (5R21AG073769).

\bibliographystyle{unsrt}
\bibliography{references}

\end{document}